# A Co-evolution Model of Network Structure and User Behavior in Online Social Networks:

## The Case of Network-Driven Content Generation

Prasanta Bhattacharya, Tuan Q. Phan, Xue Bai and Edoardo Airoldi

## Abstract

*With the rapid growth of online social network sites (SNS), it has become imperative for platform owners and online marketers to investigate what drives content production on these platforms. However, previous research has found it difficult to statistically model these factors from observational data due to the inability to separately assess the effects of network formation and network influence. In this paper, we adopt and enhance an actor-oriented continuous-time model to jointly estimate the co-evolution of the users' social network structure and their content production behavior using a Markov Chain Monte Carlo (MCMC)-based simulation approach. Specifically, we offer a method to analyze non-stationary and continuous behavior with network effects in the presence of observable and unobservable covariates, similar to what is observed in social media ecosystems. Leveraging a unique dataset from a large social network site, we apply our model to data on university students across six months to find that: 1) users tend to connect with others that have similar posting behavior, 2) however, after doing so, users tend to diverge in posting behavior, and 3) peer influences are sensitive to the strength of the posting behavior. Further, our method provides researchers and practitioners with a statistically rigorous approach to analyze network effects in observational data. These results provide insights and recommendations for SNS platforms to sustain an active and viable community.*

*Keywords: Social network structure, Content production, Co-evolution model, Homophily, Peer influence, Markov Chain Monte Carlo, Method of Moments, Latent Space Models.*

# 1 Introduction

With the proliferation of Social Network Sites (SNS), platform owners are facing increasing challenges in engaging users and, subsequently, generating revenue through advertisements (Edwards 2012; Hof 2011; Tucker 2012). Unlike other Internet-based services, the unique value of SNS lies in engaging interactions between two user roles – “content producers” who actively post, comment and share content with their friends, and “content consumers” who view and react to such content. Content producers in particular add considerable value by generating and sharing content through the network.

The content posting behavior of users as well as the users’ propensity to make new connections on social media is influenced partly by individual level factors (e.g. demographics, traits, etc.) and partly by their online social network characteristics such as their number of online friends, their network clustering, their network betweenness and so forth (Lu et al. 2013; Newman 2010). From previous research, it remains an empirical puzzle to estimate how a user’s social network, such as the number of friends or the extent of clustering in the user’s network, impacts the user’s content posting behavior. The key challenge lies in that the user’s posting behavior and social network co-evolve by affecting each other, i.e., behavior shapes the network at the same time that the network shapes behavior. From a classical social network perspective, addressing this puzzle amounts to separately assessing the effect of social influence (i.e. when network influences attitude/behavior), after controlling for any selection arising from homophily on observable or unobservable covariates (i.e. when attitude/traits influences network formation) and context effects (Borgatti and Foster 2003; McPherson et al. 2001; Shalizi and Thomas 2011). There exist certain methodological limitations with previous approaches seeking to disentangle homophily from influence. In addition, there exists a number of important theoretical gaps that remain to be addressed. For instance, earlier studies have investigated the presence of either homophily or social influence in separate contexts (Lazarsfeld et al. 1954; McPherson et al. 2001). The few that do look at their co-existence within a single context, tend to focus primarily on the relative strengths of homophily and influence (Borgatti and Foster 2003; Ennett and Bauman 1994; Kirke 2004). However, it is quite plausible that both homophily and influence play an important role in different temporal stages of the individual's life cycle, and to varying extents. In addition to this temporal dependency of homophily and influence, there could also exist a dependency on the specific state of the behavior or preference in question, i.e., is an individual equally susceptible to a change in friendship or behavior at all levels of magnitude of the behavior in question? These are critical theoretical questions that have significant practical implications for platform owners and marketers.

In the present study, we improve on previous approaches by developing an actor-based and continuous-time co-evolution model that operates under a set of Markovian assumptions to explicitly model and jointly estimate the evolution of the online social network and the evolution of online posting behavior of the user. Drawing on Snijders et al.'s (2007) framework and by leveraging prior work on latent space models (Davin et al. 2014; Hoff et al. 2002), we extend and contribute to the approach in certain key ways for SNS platforms. First, we model the co-evolution in an online dynamic behavioral setting, where the behavioral traits are not limited to a dichotomous variable, as was the case with previous studies (e.g. smoking vs. no smoking, alcoholic vs. non-alcoholic). Instead, we discretize the number of posts made by the user into quantiles. This provides us with added information about posting behavior and increased flexibility in modeling changes in behavior over time. Second, to the best of our understanding, this is the first study that attempts to adapt the actor-driven approach beyond slow-moving and relatively stable traits and behaviors (e.g. music tastes, smoking habits etc.) to a dynamic and rapidly-changing behavioral setting (e.g. online posting and messaging behavior, photo uploads etc.). Third, we correct for the presence of latent homophily based on unobservable factors, which could have potentially biased the estimates for posting influence and homophily based on similarity in posting behavior. While the presence of latent homophily has been a major confound in studies looking to disentangle influence from homophily, we exploit our longitudinal network dataset to estimate latent space positions of the actors which can potentially control for any homophily based on both observed as well as unobserved covariates (Davin et al. 2014; Goldsmith-Pinkham and Imbens 2013). Finally, prior applications of the co-evolution model have not modeled peer effects contingent on specific levels of the traits or behavior. However, we believe that individuals are likely to display varying extents of sensitivity towards peer effects depending on their current level of traits or behavior. By suitably specifying the co-evolution model, we uncover that homophily and peer influence, based on posting behavior, are sensitive to the current level of the users' posting behavior.

Our method has advantages over existing methods in several key aspects. First, prior experimental approaches to addressing similar research questions usually estimate either influence or homophily while controlling for the other (Sacerdote 2001; Toubia and Stephen 2013). Our technique offers an improvement over these methods in that we explicitly model and jointly estimate both friendship formation as well as social influence. Further, in using field data from a real world context, we also achieve a higher ecological validity. Second, previously used non-experimental approaches like the contingency table approach (Fisher and Bauman 1988; Kandel 1978), the aggregated network based method (Kirke 2004; Yoganarasimhan 2012), or the structural equation models (Iannotti et al. 1996; Krohn et al. 1996) tend to either ignore the network dependence of users (i.e. assume dyadic independence) or fail to take into account the possibility of errors introduced due to incomplete observations, as is often the case with such discrete-time models. Our method, in comparison, adequately addresses both these challenges by employing a continuous-time structural model that makes no assumptions of dyadic independence.

The estimation of network-behavior co-evolution models is often non-trivial. Closed form solutions are generally not possible for the likelihood function in such models, making estimation methods such as maximum likelihood or Bayesian estimation inadequate. To overcome this hurdle, we resort to a simulation-based estimation framework based on Markov Chain Monte Carlo (MCMC) estimations (Snijders 2001; Steglich et al. 2010). Specifically, we use a MCMC-based Method of Moments (MoM) estimator to estimate the co-evolution parameters in our model. While some prior studies have used computational simulations to model endogenous evolution of network ties and individual attributes (Carley 1991; Macy et al. 2003), our model allows for statistical inference testing, model fit assessments and counterfactual simulations. Moreover, it allows for different forms of objective functions and operates under an acceptable set of assumptions (e.g., conditional independence etc.).

Using this approach, we can make the following inferences about the nature and extent of peer effects as well as homophilous peer selection in content production on SNS. First, we find evidence for homophily based on similarity in content posting behavior, but not on individual-level covariates, like age or gender. Second, we observe the existence of peer influence, but in a direction opposite to that of homophilous interaction. We find opposing roles of behavioral similarity at different stages of friendship formation. Surprisingly, however, individuals befriend others who are similar in content production behavior during the friendship formation stage, but gradually diverge others over time afterwards. Third, we provide evidence that the strength of homophilous friend selection as well as social influence varies as a function of the specific level of the behavior. Specifically, we find that low content posters are more susceptible than heavy content posters to homophilous friend selection. However, once they make friends, low posters are more likely to diverge from their peers as compared to heavy posters. Moreover, we show that these findings are robust to the presence of potential latent homophily arising from unobservable factors.

In summary, our study offers a statistically disciplined approach to modeling the co-evolution of online social network structure and posting behavior on the SNS. Using and extending a MCMC-based stochastic simulation model, we uncover insights about the mechanisms that drive peer effects and the behavioral dependency of these mechanisms on SNS. The findings from this study can also be used to generate actionable recommendations for social network platform owners, social media marketers and advertisers. For example, our results can inform and guide the design of better and more adaptable friendship recommendation engines for SNS platforms, while also informing social media marketers on how to effectively seed marketing information, and target valuable users on the SNS.

In the following section, we present a summary of previous studies that discuss peer effects in social networks and a relatively newer set of studies that have used the co-evolution model in varying contexts. Next, we offer a brief summary of the co-evolution model that we use in our empirical analyses. Following this, we discuss our empirical setting and demonstrate our findings. We conclude with a discussion of the key contributions of our study, the limitations, and a roadmap for future research.

# 2 Related Work

## 2.1 Peer effects in social networks

Social science researchers have always been interested in understanding the interdependence between the behavior of group members and the group's structure, as reflected by the inter-member ties within the group. For instance, sociologists and psychologists have long discussed the effect of social cohesion among group members on norm compliance and deviance (Asch 1951; Durkheim 1884; Homans 1961). Researchers have also investigated the role of individual actions on emergent social outcomes and social structures (Emirbayer and Goodwin 1994; Homans 1961; Stokman and Doreian 1997).

More recently, researchers have observed that the preference and behavior of individuals tend to be more similar when they are connected in a relationship, than when they are not (Hollingshead 1949; Newcomb 1962). This phenomenon has been studied under various names, the most common of which are homogeneity bias (Fararo and Sunshine 1964) and network autocorrelation (Doreian 1989). The increased focus on understanding the mechanisms that lead to such network autocorrelation was initially driven by a need to understand the onset, and diffusion of addictive behaviors, including smoking, alcoholism, and substance-abuse among adolescents (Brook et al. 1983; Cohen 1977; Kandel 1978). However, over time, network autocorrelation has been also observed and studied extensively in the context of online social networks (Aral and Walker 2014; Aral et al. 2009, 2013; Lewis et al. 2012).

While some sociologists and social psychologists propose the idea of social influence or network-driven assimilation as a potential cause of such effects (Asch 1951; Friedkin 2001; Oetting and Donnermeyer 1998; Singh and Phelps 2013), others propose selection-based mechanisms like homophily to explain why such effects might occur (Aral et al. 2009, 2013; Lazarsfeld et al. 1954; McPherson et al. 2001; Nahon and Hemsley 2014). A third line of research challenges both influence and homophily based explanations and, instead, focuses on the role of a shared context between the networked individuals as a driving factor (Feld and Elmore 1982; Feld 1981, 1982). Borgatti and Foster (2003)described these competing perspectives in terms of the temporal ordering and causal validity of network or behavioral change (Borgatti and Foster 2003). Specifically, they suggest that if behavior is the consequence of network change, then this is explained by peer influence. If, however, the network is the consequence of behavior change, then this is explained by selection mechanisms such as homophily, but only if the temporal antecedence is causal. If it is merely correlational, then this can also be explained as a result of shared social contexts. Understanding what drives network autocorrelation in various contexts remains an open empirical puzzle, and several previous works have pointed out this underlying tension among the competing perspectives (Ennett and Bauman 1994; Kirke 2004; Michell and Pearson 2000; Pearson and West 2003).

It is a challenging exercise to disentangle these competing mechanisms in real-world contexts. While there have been several different approaches, there has been limited progress in identifying competing mechanisms through using observable data. Experimental approaches using lab and field studies have attempted to intervene with either the network or the behavior in order to identify the other (Aral and Walker 2014; Asch 1951; Herman et al. 2003; Sacerdote 2001). These approaches have been useful in uncovering causal relationships, but often at a price of reduced ecological validity. Also, it is fairly simple to imagine contexts where such methods would be untenable. For instance, it is debatable whether friendship formation can truly be exogenized without losing realism. In addition, while experimental approaches are considered to be the holy grail of inference testing, some can be difficult to execute, while others face issues with ecological and external validity of the population (Berkowitz and Donnerstein 1982; Falk and Heckman 2009). Moreover, extensive longitudinal field studies are complicated to design, time consuming, loosely controlled, and potentially face human subject regulations. These challenges have limited the applicability of experiments in network research.

As a result, others have attempted to uncover the impact of peer effects using non-experimental methods. Such attempts can be largely classified under three major categories. The first is the contingency table approach (Billy and Udry 1985; Fisher and Bauman 1988; Kandel 1978), in which dyads of mutually selected friends are selected and cross-tabulated across subsequent periods. The observed measures on a behavioral attribute are similarly recorded. Estimate for influence is then obtained from pairs of individuals whose friendship is preserved over subsequent periods, but who show a change in behavior. In a similar fashion, estimates for selection are assessed from pairs of individuals whose friendship ties change over subsequent periods, but show identical behavior in both periods. A second approach is the aggregated personal network approach which follows a two-step strategy (Cohen 1977; Davin et al. 2014; Ennett and Bauman 1994; Kirke 2004; Pearson and West 2003). In the first step, the user's network characteristics (e.g. network structure measures) are collapsed into individual measures (e.g. user's transitivity, betweenness centrality, etc.). In the second step, these measures are used to predict user-level outcomes under an implicit assumption that such measures are independent across observations. Finally, the structural equation modeling approach attempts to model the cross-lagged panel of latent network-level constructs (Iannotti et al. 1996; Krohn et al. 1996). This particular approach is better than the previous two approaches by virtue of its ease of modeling and estimating selection and influence effects simultaneously using a system of equations.

There are three major shortcomings with all the above mentioned techniques for studying peer effects using longitudinal network data. First, most prior methods tend to ignore the network dependence of users. Thus, the assumption of independence across observations is clearly violated in such settings. Second, these methods tend to control or even ignore alternate mechanisms of network or behavior evolution such as the impact of shared social contexts. Lastly, these methods are problematic in the presence of incomplete observations, as is often the case with longitudinal discrete-time datasets where observations about the user and the network are only made at specific points in time, with little information about the inter-period dynamics. However, ignoring the evolutionary dynamics between discrete time periods can significantly affect our ability to make inferences about peer effects, as pointed out by Steglich et al. (2010).

## 2.2 Content production in online social networks

Social network sites (SNS) have been a subject of active research in several disciplines including information systems, marketing, social psychology and computer science. boyd and Ellison (2007) define SNS as “web-based services that allow individuals to (1) construct a public or semi-public profile within a bounded system, (2) articulate a list of other users with whom they share a connection, and (3) view and traverse their list of connections and those made by others within the system.” While there have been extensive studies looking at how prolonged use of SNS influences the psychological well-being of the users and how this process of generating online social capital differs from offline users (Hargittai 2007; Steinfield et al. 2008; Valkenburg et al. 2006; Wellman et al. 2001), others have taken a more normative approach to discuss how user engagement increases on the SNS (Fogg and Eckles 2007) and whether engagement on SNS has a positive or negative impact on its users (Binder et al. 2009; Livingstone 2008). Finally, there have been some exemplary efforts probing how organizations use social media to engage more effectively with their target users both inside and outside the organization (Sinclaire and Vogus 2011; Steinfield et al. 2009; Waters et al. 2009).

The creation and spread of user generated content (UGC) as a means of online word-of-mouth (WOM) has interested social network researchers for several decades, and has been extensively used by brand marketers to understand and increase brand awareness, evaluation and sales (Chevalier and Mayzlin 2006; Goh et al. 2013; Reingen 1984). The common thread that emerges from the extant literature on WOM is the value of WOM behavior for both the users as well as the platform owners; Higher WOM levels essentially lead to higher engagement levels on the platform. We observe that high levels of self-disclosure on online platforms allow the sites to collect essential user data that can then be used in marketing implementations. Further, ensuring a persistent and critical mass of users on the platform enables advertisers to monetize by delivering more advertisements to the users in a targeted fashion (Goldfarb and Tucker 2011).

A number of studies have investigated the reasons why users develop a propensity to contribute public content. Using a natural experiment, Zhang and Zhu (2011) talk about public contributions on Chinese Wikipedia and show that the users are often motivated to make contributions to public goods in anticipation of social benefits and intrinsic "warm glow" effects, in addition to the marginal utility which can be obtained by provisioning of the public content (Andreoni 1989; Konow 2006; Toubia and Stephen 2013). Furthermore, the uses and gratifications theory has been successfully extended to understand the motivations behind Internet use (LaRose and Eastin 2004) as well as the motivations guiding the use of SNS groups (Park et al. 2009). These studies have found that SNS provide distinct gratifications through communication, entertainment, information, and status seeking.

Previous work using historical data, however, faces limitations by ignoring the interdependencies of the underlying network structure and the content production process. Consequently, there has been very little work that looks at objectively investigating and solving the network autocorrelation problem in online contexts (Backstrom et al. 2006; Crandall et al. 2008; Singla and Richardson 2008). The few attempts that exist focus primarily on establishing the presence of either influence or homophily and do not provide a flexible model that is geared towards performing stronger inference testing. A few exceptions to this are the recent studies by Aral et al. (2009) and Snijders et al. (2007). Both these models follow fundamentally different approaches for trying to separate homophily from influence. While Aral et al. (2009) use a matched sample estimation framework that hinges on the presence of several user-specific attributes and preferences to perform suitable matching, Snijders et al.(2007) use a more parsimonious random-graph based model in an offline setting with relatively stable behaviors like smoking and alcohol consumption. In the following section, we describe and extend on Snijder's approach and illustrate the utility of this model in disentangling social effects for dynamic and non-stationary behavior in an online setting.

### 2.3 Stochastic actor-driven co-evolution model

The co-evolution models offer a continuous time scenario in which users simultaneously alter their network ties as well as their behavior at random instants in time, which may or may not be observed by the researcher. Similar Markovian models for longitudinal social network data have a rich history of use in the social networks literature (Holland and Leinhardt 1977; Wasserman 1977). Such continuous time models, in principle, provide greater flexibility and theoretical grounding than comparable discrete time models (Katz and Proctor 1959; Wasserman and Iacobucci 1988). However, some of the earlier continuous-time Markov chain models, like the reciprocity model (Wasserman 1977, 1980a), possess two main limitations. First, the models assume dyadic independence in the social network, which makes the analysis computationally convenient, but is untenable in most real-world contexts. Second, such models face restricted capability with parameter estimation and subsequent counterfactual analyses (Mayer 1984; Wasserman 1980b).

The above limitations were largely mitigated by the use of Monte Carlo Markov Chain (MCMC) based stochastic simulation models for sociometric data, as proposed by Snijders (1996) and later extended empirically in de Bunt et al. (1999). However, these models dealt with the issue of network evolution without focusing on any associated behavior. In a later work, Snijders et al. (2007) extended this actor-driven network evolution model to explicitly model the interrelationships between the network and the user's behaviors, and applied it to smoking and alcohol consumption behaviors. This new framework analyzed the network and set of user behaviors together in a joint state space and modeled how the network and behaviors evolved by influencing each other. The model accounted for network dependence of users and allowed researchers to investigate any number of alternative mechanisms of peer effects. A number of recent studies have used this co-evolution model to investigate the effects of selection and influence on social behaviors such as substance abuse among friends (Steglich et al. 2010), diffusion of innovation (Greenan 2015) as well as the evolution of self-reported music and movie tastes among adolescents (Lewis et al. 2012).

In our study, we develop a co-evolution model for large-scale observational data on the network and posting-behavior of SNS users. Our model sets out to investigate how peer effects influence dynamic content production (e.g. posting public content) on the SNS. Unlike previous studies that have investigated co-evolution of network and behavior in an offline context with self-reported network and behavior data, the current research uses objective network and posting behavior data from a large social network site, the largest SNS in the world. Moreover, while previous studies have predominantly focused on stable behaviors like smoking and alcoholism, which do not change frequently over time, the current study focuses on dynamic behaviors, like content production which has a higher frequency of change. Furthermore, we extend the previous methods to model non-binary behaviors by discretizing online posting behavior based on several quantiles of intensity (e.g. ranging from levels 1 to 10). We posit that understanding the evolution of such online behavior is valuable to platform owners, marketers, and advertisers. However, due to the fast-changing nature of the behavior, it has been increasingly difficult for existing discrete-time models to accurately capture and predict these dynamics. In the following section, we describe the co-evolution model used in our setting and illustrate how we perform parameter estimation of the network and behavior effects.

# 3 Co-evolution Model of Networks and Behavior

We develop an actor-based continuous-time model for the co-evolution of online network formation and content generation. Our model builds upon and extends Snijders et al. (2007) and Steglich et al. (2010) in several key ways and is applied to a unique panel dataset obtained from a large social network site, the largest online social network in the world. This network-behavior co-evolution model draws upon past work on actor-oriented pure network evolution models (Snijders 2001).

## 3.1 The model

We observe a network with $N$ users, for a total of $T$ months, and model two main variables, namely, the state of the time-varying friendship network, a $N \times N$ matrix $A_t$ and a $N \times 1$ time-varying integer-valued posting behavior vector $P_t$, which denotes the number of public posts contributed by users at time $t$.

### 3.1.1 Timing of decision

We assume that the evolution of both the network as well as the behavior follows a first-order Markov process, using very small time-increments, called "micro-steps" that occur at random instants in time. The network evolves in continuous-time but is observed at discrete moments. At a given micro-step, we constrain the network or the behavior to only allow a unit change, i.e., a tie forms or dissolves, or the posting volume increases or decreases by 1 unit. Using a Poisson process, we model these specific points in time when any given user $i$ gets the opportunity to make a decision to change the vector of her outgoing tie variables $a_{ij} = [A]_{ij}$, $j = 1, \dots N - 1$, or her behavior variable $p_i = [P]_i$.

Consequently, the rates at which the users make network decisions ($\lambda_i^{[A]}$) and behavioral decisions ($\lambda_i^{[P]}$) between time periods $t$ and $t + 1$ are decided by rate functions as described in Eqs. 1and 2 below.

$$\lambda_i^{[A]}(A_t, P_t) = \rho_m^{[A]} exp^{(h_i^{[A]}(\alpha^{[A]}, A_t, P_t))} \quad (network\ decisions) \tag{1}$$

$$\lambda_i^{[P]}(A_t, P_t) = \rho_m^{[P]} exp^{(h_i^{[P]}(\alpha^{[P]}, A_t, P_t))} \ (behavioral\ decisions) \qquad (2)$$

where, the parameters $\rho_m^{[A]}$ and $\rho_m^{[P]}$ are dependent on the observed discrete time-period and capture periodic variations in either network or posting behavior[1], and the functions $h_i^{[A]}(.)$ and $h_i^{[P]}(.)$ model dependence on the current state of the network and the posting behavior. The exact functional forms of $h_i^{[A]}(.)$ and $h_i^{[P]}(.)$ depend on the network and behavioral effects that we choose to model in our context, and we specify these in detail in Sec. 3.2. However, in the current model specification, we assume that the rate functions are constant across the actors and are only dependent on the specific discrete observation periods *m*.

#### 3.1.2 Objective function

While the rate functions model the timing of the users' decisions (i.e. to change network or behavior), the objective functions model the specific changes that are made. A user *i* optimizes an objective function in the current time period over the set of feasible micro-steps she can take. This objective function is composed of three parts (Steglich et al. 2010): the evaluation functions $f_i^{[A]}$ and $f_i^{[P]}$, the endowment functions $g_i^{[A]}$ and $g_i^{[P]}$, and random disturbances $\epsilon_i^{[A]}$and $\epsilon_i^{[P]}$, capturing residual noise.

The evaluation functions are parameterized by the vectors $\beta^{[A]}$ and $\beta^{[P]}$; the endowment functions are parameterized by the vectors $\gamma^{[A]}$ and $\gamma^{[P]}$ , as shown in Eqs. (3) and (4) below.

---

[1] Estimating the rate functions $\lambda_i^{[A]}(.)$ and $\lambda_i^{[P]}(.)$ is similar to computing the ratio of network and behavior changes respectively in period m, to the total number of network and behavior changes across all m. However, the reason the parameters $\rho_m^{[A]}$ and $\rho_m^{[P]}$ are estimated from data and not just computed as a ratio is because this ignores the possibility of the actor not changing her network/behavior (or even reverting it). Consequently, the estimated rate functions will always be higher than the observed average number of changes.

$$f_i^{[A]}(\beta^{[A]}, A_t, P_t) + g_i^{[A]}(\gamma^{[A]}, A_t, P_t | A_{t-1}, P_{t-1}) + \epsilon_i^{[A]}(A_t, P_t) \ (network\ decisions) \quad (3)$$

$$f_i^{[P]}(\beta^{[P]}, A_t, P_t) + g_i^{[P]}(\gamma^{[P]}, A_t, P_t | A_{t-1}, P_{t-1}) + \epsilon_i^{[P]}(A_t, P_t) \ (behavioral\ decisions) \quad (4)$$

The evaluation functions capture the utility obtained by a user $i$ from her network-behavior configuration. The functions $f_i^{[A]}(.)$ and $f_i^{[P]}(.)$ in Eqs. (3) and (4) provide a measure of fitness of the state of the network and posting behavior, as perceived by the users. This implies that users constantly strive to make specific changes to their friendship network and posting behavior to maximize the value of this evaluation function.

The endowment functions $g_i^{[A]}(.)$ and $g_i^{[P]}(.)$, from (3) and (4) above, capture the part of utility that is lost when either the network ties or the posting behavior is changed by a single unit, but which was obtained without any "cost" when this unit was gained earlier. In other words, such endowment functions are useful to model situations where the creation and dissolution of ties, or an increase or decrease in posting behavior are asymmetric in terms of utility gained or lost. However, since in the context of our study, we do not model deletion of friends on the platform or the deletion of content, we do not include such endowment functions in our model.

### 3.1.3 Choice probabilities and intensity matrix

The final term in the objective function described in (3) and (4) above are the set of random and i.i.d. residuals $\epsilon_i^{[A]}$ and $\epsilon_i^{[P]}$. As is the case with random utility models, if we assume that these residuals follow type-1 extreme value distribution, it allows us to write the resulting choice probabilities for the network and posting micro-step decisions as a multinomial logit (Maddala 1986). For the network micro-step decision, the resulting choice probability is illustrated in (5) below.

$$Pr\left(a_{t+1} = a_t + \delta \middle| a_t, p_t, \beta^{[A]}\right) = \frac{\exp(f_i^{[A]}(\beta^{[A]}, a_{t+1} = a_t + \delta, p_t))}{\sum_{\varphi} \exp(f_i^{[A]}(\beta^{[A]}, a_{t+1} = a_t + \varphi, p_t))} \tag{5}$$

where, $a_{t+1}$ is the resulting network at $t+1$ when a user $i$ at micro-step $t$ alters the value of her tie variables by $\delta$ (or $\varphi$) where, $\delta, \varphi \in \{0,1\}$, i.e., user $i$ either creates a new tie or makes no change to her network[2]. Similarly, for the posting micro-step decision, the resulting choice probability is illustrated in (6) below.

$$Pr\left(p_{t+1} = p_t + \delta \middle| a_t, p_t, \beta^{[P]}\right) = \frac{\exp(f_i^{[P]}(\beta^{[P]}, a_t, p_{t+1} = p_t + \delta))}{\sum_{\varphi} \exp(f_i^{[P]}(\beta^{[P]}, a_t, p_{t+1} = p_t + \varphi))} \tag{6}$$

where, $p_{t+1}$ denotes the resulting state of posting behavior in $t+1$ when user $i$ changes her posting volume at micro-step $t$ by a factor of $\delta$ (or $\varphi$), where, $\delta, \varphi \in \{-1, 0,1\}$ i.e. the user increases her positing volume by 1 unit, decreases it by 1 unit or makes no new posts.

Once we have formulated the choice probabilities, the subsequent transition matrix Q, also called as the intensity matrix, models the transition from state $(a_t, p_t)$ at micro-step $t$ to a new state $(a_{t+1}, p_{t+1})$ at micro-step $t+1$, and can be specified by the following entries.

$$Q(a_{t+1}, p_{t+1}) \tag{7}$$

$$= \begin{cases} \lambda_i^{[A]} Pr(a_{t+1} = a_t + \delta | a_t, p_t), \text{if } (a_{t+1}, p_{t+1}) = (a_t(i,\delta), p_t); \\ \lambda_i^{[P]} Pr(p_{t+1} = p_t + \delta | a_t, p_t) \text{ if } (a_{t+1}, p_{t+1}) = \left(a_t, p_t(i,\delta)\right); \\ -\sum_i \left\{ \sum_{\delta \in \{-1.1\}} Q(a_t(i,\delta), p_t) + \sum_{\delta \in \{-1.1\}} Q\left(a_t, p_t(i,\delta)\right) \right\}, \text{if} (a_{t+1}, p_{t+1}) = (a_t, p_t); \text{ and} \\ 0, \text{ otherwise.} \end{cases}$$

[2]There is no observed case of friendship dissolution (i.e. 1 to 0) in our data context.

### 3.1.4 Model estimation

Due to the complexity of explicitly computing the likelihood function, we employ the use of simulation-based estimators. Specifically, we use a Markov Chain Monte Carlo (MCMC) based Method-of-Moments (MoM) estimator to recover the parameters of these rate and evaluation functions. The MoM estimator for our data and the parameters is based on the set of network and behavior related statistics that are specified in the following section. The MCMC implementation of the MoM estimator uses a stochastic approximation algorithm that is a variant of the Robbins-Monro (1951) algorithm (Robbins and Monro 1951) as detailed in Appendix 1.

The following section describes the empirical context for testing 1) the proposed co-evolution model to investigate the presence of peer effects and 2) the dependence of these peer effects on the state of the posting behavior.

## 3.2 Model parameterization in context of SNS

In our context, the functions $h_i^{[A]}(.)$, $h_i^{[P]}(.)$, $f_i^{[A]}(.)$ and $f_i^{[P]}(.)$ from (1), (2), (3) and (4) can be modelled as a weighted sum of various network characteristics (e.g. degree, transitivity, homophily based on user covariates etc.) and behavioral characteristics (e.g. behavior trends, similarity measure, effect of user covariates on behavior etc.). We denote the matrix of network and behavior statistics computed in each time period t by $\mathrm{S}_\mathrm{t}^{[\mathrm{A}]}$ and $\mathrm{S}_\mathrm{t}^{[\mathrm{P}]}$, which are $\mathrm{N} \times \mathrm{K}_1$ and $\mathrm{N} \times \mathrm{K}_2$ matrices of $\mathrm{K}_1$ network and $\mathrm{K}_2$ behavioral characteristics, respectively. The functions $\mathrm{h}_\mathrm{i}^{[\mathrm{A}]}(.)$ and $\mathrm{h}_\mathrm{i}^{[\mathrm{P}]}(.)$ from the rate functions are specified as follows.

$$h_i^{[A]}\left(\alpha^{[A]}, A_t, P_t\right) = \sum_q \alpha_q^{[A]} \, s_{iqt}^{[A]}(A, P) \tag{8}$$

$$h_i^{[P]}\left(\alpha^{[P]}, A_t, P_t\right) = \sum_r \alpha_r^{[P]} s_{irt}^{[P]}(A, P) \quad (9)$$

Here, $\alpha_q$ indicates dependence on the statistics $s_{iqt}^{[A]}(A, P)$, and q $\subset K_1$. Similarly, coefficient $\alpha_r$ indicates dependence on the statistics $s_{irt}^{[P]}(A, P)$, and r $\subset K_2$, where $s_{iqt}^{[A]}(A, P))$ and $s_{irt}^{[P]}(A, P))$ are vectors of one-dimensional statistics defined for each user i, and used to capture the rate dependence on the user's network characteristics (e.g. out-degree) and behavioral characteristics (e.g. SNS tenure) respectively. For the current set of analyses, however, we hold both sets of rate functions to be constant across all actors, and model only the dependence on the time period i.e. parameters $\rho_m^{[A]}$and $\rho_m^{[A]}$ in Eqs. 1 and 2.

Similarly, the functions $f_i^{[A]}(.)$ and $f_i^{[P]}(.)$ can be specified follows.

$$f_i^{[A]}\left(\beta^{[A]}, A_t, P_t\right) = \sum_{k_1} \beta_{k_1}^{[A]} s_{ik_1}^{[A]}(A_t, P_t) \quad (network\ evaluations) \quad (10)$$

$$f_i^{[P]}\left(\beta^{[P]}, A_t, P_t\right) = \sum_{k_2} \beta_{k_2}^{[P]} s_{ik_2}^{[P]}(A_t, P_t) \quad (behavior\ evaluations) \quad (11)$$

where, $s_{ik_1}^{[A]} = [S^{[A]}]_{ik_1}$ is the ${k_1}^{th}$ network statistic of user $i$, and, similarly, $s_{ik_2}^{[P]} = [S^{[P]}]_{ik_2}$ is the ${k_2}^{th}$ behavioral statistic of user $i$.

We parameterize the objective function based on our current research context, that of online posting behavior among a student population on a large and popular SNS. Specifically, we seek to investigate the presence of homophilous friendship formation based on similarities in posting behavior, as well as the role of peer influence in regulating content generation over time. Furthermore, we also analyze the dependency of peer effects on the specific state of the posting behavior to investigate whether active content posters react differently to peer effects as compared to less active posters.

### 3.2.1 The presence of homophily and peer influence

In this section, we define and specify estimation statistics for both the network as well as the posting behavior effects which we model in our study.

***Social network effects***

The network effects from $S_t^{[A]}$ that we model are the user *i*'s out-degree ($s_{i1t}^{[A]}$), the transitivity ($s_{i2t}^{[A]}$), homophily effects based onposting behavior($s_{i3t}^{[A]}$), and homophily based on the covariates, gender ($s_{i4t}^{[A]}$), age($s_{i5t}^{[A]}$), and SNS tenure ($s_{i6t}^{[A]}$). We also include effects that model the influence of individual covariates i.e., gender ($\mathrm{Gender}_i^{[A]}$), age $\left(\mathrm{Age}_i^{[A]}\right)$ and social network site (SNS) tenure ($\mathrm{SNS\ Tenure}_i^{[A]}$), on the propensity to form new friends. The mathematical illustrations are provided in Eqs.12 through 17.

*(i) Degree ($s_{i1t}^{[A]}$) and Transitivity ($s_{i2t}^{[A]}$)*

$$s_{i1t}^{[A]}(a) = \sum_j a_{ijt} \qquad (12)$$

$$s_{i2t}^{[A]}(a) = \sum_{j,h} a_{ijt} * a_{jht} * a_{iht} \qquad (13)$$

*(ii) Homophily based on posting behavior and covariates (gender, age, SNS tenure)*

$$s_{i3t}^{[A]}(a, p) = a_{i+t}^{-1} \sum_j a_{ijt} \left(1 - \frac{|p_{it} - p_{jt}|}{R_{pt}}\right) \qquad (14)$$

where, $R_{pt}$ is the range of the posting variable *P* at step *t*. Variable $s_{i3t}^{[A]}$ represents the effect of homophily, based on posting behavior, such that $s_{i3t}^{[A]}$ takes a higher value for those users whose posting volume is closer to that of their peers (i.e. the value of $|p_{it} - p_{jt}|$ is small). Thus, a drive towards a higher value of $s_{i3t}^{[A]}$ can be seen as an increased propensity towards creating homophilous friendships based on similarity in posting behavior.

For covariates $X = \{\text{gender}, \text{age}, \text{SNStenure}\}$, we have similar expressions for $s_{i4t}$, $s_{i5t}$ and $s_{i6t}$ respectively.

*(iii) Covariate ($X_j$) on the Degree effects (i.e., effect of user's gender, age, and SNS tenure on her Degree)*

$$\text{Gender}_{it}^{[A]}(a, x) = \sum_j a_{ijt} * x_{1i} \tag{15}$$

$$\text{Age}_{it}^{[A]}(a, x) = \sum_j a_{ijt} * x_{2it} \tag{16}$$

$$\text{SNSTenure}_{it}^{[A]}(a, x) = \sum_j a_{ijt} * x_{3it} \tag{17}$$

$\text{Gender}_{it}^{[A]}$ represents the effect of the user $i$'s gender ($x_1$) on her propensity to make new friends during step *t*, such that a positive and significant estimate on the statistic would imply that females *(Gender = 1)* make more friends than males *(Gender = 0)*, and vice versa. We have similar expressions for $\text{Age}_{it}^{[A]}$ and $\text{SNS Tenure}_{it}^{[A]}$ respectively. In all the above equations, $a_{ijt} = 1$ if a tie exists between *i* and *j* in step *t*, and 0 otherwise.

### *Posting behavior effects*

Next, we specify the rate and evaluation functions as defined for the posting behavior. In Eqs.11, the behavior effects that we model are the user's behavior tendency effect ($s_{i1t}^{[P]}$), the peer influence effect i.e. social influence $s_{i2t}^{[P]}$, and effects that capture the influence of individual covariates like gender ($\text{Gender}_{it}^{[P]}$), age ($\text{Age}_{it}^{[P]}$) and SNS tenure ($\text{SNS Tenure}_{it}^{[P]}$) on the posting behavior, *P*. We provide the mathematical illustrations in (18) through (22).

*(i) Behavioral tendency effect (This captures the natural tendency of users to increase or decrease behavior over time)*

$$s_{i1t}^{[P]}(a, p) = p_{it} \tag{18}$$

*(ii) Peer influence effect (The propensity of users to assimilate in behavior towards their peers)*

$$s_{i2t}^{[P]}(a, p) = a_{i+t}^{-1} \sum_j a_{ijt} \left(1 - \frac{|p_{it} - p_{jt}|}{R_{pt}}\right) \tag{19}$$

where, $R_{pt}$ is the range of the posting variable P. $s_{i2t}^{[P]}$ represents the effect of peer influence, based on posting behavior, such that $s_{i2t}^{[P]}$ would have a higher value for those users whose posting volume is closer to that of their peers (i.e. the value of $|p_{it} - p_{jt}|$ is smaller). Thus, a positive and significant estimate on this statistic would indicate that users regulate their posting behavior to assimilate with their peers i.e. matching the posting rate of peers, and vice versa.

*(iii) Influence of covariates (i.e. gender, age, SNS tenure) on behavior*

$$\text{Gender}_{it}^{[P]}(p, x) = p_{it} * x_{1i} \tag{20}$$

$$\text{Age}_{it}^{[P]}(p, x) = p_{it} * x_{2it} \tag{21}$$

$$\text{SNS Tenure}_{it}^{[P]}(p, x) = p_{it} * x_{3it} \tag{22}$$

Here, $\text{Gender}_{it}^{[P]}$ represents the effect of gender ($x_{i1t}$) on posting behavior such that a significant and positive estimate on this statistic would indicate that females *(Gender = 1)* post more than males *(Gender = 0)*. Eqs. 21 and 22 denote similar expressions that represent the effects of age and SNS tenure on posting behavior respectively.

It is clear from the above formulation of effects, that the mathematical illustration for the network and behavior effects to compute homophily (Eq. 14) and peer influence (Eq. 19) are identical. This point lies at the core of the problem that is separating the effect of homophilous selection from peer influence. However, we exploit the longitudinal nature of our dataset to successfully identify temporal sequentiality across the periods. In other words, we use dyads of users who first become friends and then converge in behavior, to identify influence. Similarly, we use dyads of users who show similarity in behavior before becoming friends, to identify homophily. While there might be other latent confounds that we do not capture in our modeling, our approach makes an attempt at demonstrating a restricted form of causality. This view is consistent with several recent studies investigating related topics on homophily and influence among student populations (Lewis et al. 2012; Steglich et al. 2010).

### 3.2.2 Behavioral dependency of homophily and peer influence

While homophilous or assortative relationships among individuals have been reported extensively in previous research on the subject (Aral et al. 2009; McPherson et al. 2001; Park and Barabási 2007), what remains to be investigated is whether such homophilous selection effects vary in strength depending on the current state of the observable attribute or behavior. For instance, consider how an individual who smokes cigarettes is more likely to make friends with a fellow smoker (Christakis and Fowler 2008; Pearson and West 2003). However, would his affinity to make friends with a similar smoker be any higher or lower depending on how many cigarettes he smokes each day at the present moment? An analogous problem arises in studying influence. It has been widely observed that peer influence plays an important role in the onset and sustenance of various addictive behavior, including smoking (Christakis and Fowler 2008; Ennett and Bauman 1994). However, little is known about whether such peer influence effects are particularly stronger or weaker for different levels of the behavior itself.

In our study, we investigate whether SNS users show varying strengths of selection bias due to homophily and susceptibility to peer influence depending on their current levels of posting behavior. To achieve this, we cluster all users depending on their levels of posting-behavior into three major categories. Based on the volume of content generated, we categorize the top 10 percentile of individuals in each time period as Most Active Posters (MAP), and categorize the bottom 10 percentile of individuals as Least Active Posters (LAP). All other users are categorized as Moderately Active Posters (MoAP). We introduce dummy variables for each of the first two groups in our model, keeping the middle group as our baseline. This is shown in Eqs. 23 and 24 below. The estimates from the interaction between these dummy variables and our homophily and peer influence variables would help us address our question at hand.

$$s_{i7t}^{[A]}(a,p) = \mathrm{MAP_i} * a_{i+t}^{-1} \sum_j a_{ijt} \left(1 - \frac{|p_{it} - p_{jt}|}{R_{pt}}\right), \text{and} \tag{23}$$

$$s_{i8t}^{[A]}(a,p) = \mathrm{LAP_i} * a_{i+t}^{-1} \sum_j a_{ijt} \left(1 - \frac{|p_{it} - p_{jt}|}{R_{pt}}\right) \tag{24}$$

where, $R_{pt}$ is the range of the variable $P$.In the above equations, the $\mathrm{MAP_i}$ and $\mathrm{LAP_i}$ dummy variables denote whether a user *i* is a heavy poster or low poster. The middle group ($\mathrm{MoAP_i}$) is held as the baseline group for comparison of estimates. Similar effects are constructed for the interaction of these activity dummies and the behavioral homophily effect ($s_{i3t}^{[P]}$ and $s_{i4t}^{[P]}$).

# 4 Estimation Results

## 4.1 Data context

We obtained complete online network data from a large social network site for 2507 undergraduate students attending a North American university for the months from September 2008 till February 2009. Additionally, we recorded the number of monthly public posts made by these users on the social media platform during the same period. The descriptive statistics of the key variables are illustrated in Table 1 below.

| | Min | Max | Mean | Std. Dev. |
|---|---|---|---|---|
| **Dependent Variable:** | | | | |
| Total Monthly Public Posts ($P_{it}$) | 0.000 | 15200.360 | 145.148 | 492.811 |
| **Independent Variables:** | | | | |
| Biological Age ($Age_i$) [years] | 20.000 | 26.000 | 22.244 | 1.381 |
| SNS Tenure, (SNS $Tenure_i$) [days] | 831.000 | 2591.000 | 1778.376 | 344.891 |
| Gender ($Gender_i$) | 0.000 | 4.000 | 1.471 | 0.543 |
| Number of friends added on SNS ($D_{it}$) | 1.000 | 98.000 | 6.270 | 7.055 |
| Total Monthly Public Posts by Friends ($\sum_j P_{jt-1}$) | 0.000 | 1293232.000 | 47006.730 | 79376.770 |

**Table 1 Descriptive summary of model variables**

$P_{it}$ and $D_{it}$ constitute the key variables for the co-evolution model, depicting the total number of monthly public posts and new friends added on the SNS respectively. The covariates include $Gender_i$, the gender of the user, $Age_i$, the biological age of the user, and $SNS\ Tenure_i$, the total number of days spent by the user on the SNS at the time of recording the data.

The network and behavior descriptive summaries are detailed in Appendices 2 and 3. Within our observation period, the students produced a substantial amount of content on the social media platform, and also established several new friendships. This provides us with sufficient variability in our data to test our proposed models.

## 4.2 The evolution of homophily and peer influence

We estimate the rate and evaluation functions from the co-evolution model as specified in Sections 3.1 and 3.2.1 earlier using a Method of Moments (MoM) estimator and present the results in Table 2. The MoM estimator essentially tries to recover parameter estimates by matching the observed network data with the simulated network data. Appendices 4 and 5 provide details on the convergence descriptives for these simulations. Specifically, we provide information about the deviation of our simulated network and behavioral statistics from the observed data. Tables 2(a) and 2(b) highlight the estimation results for rate parameters $\rho_m^{[A]}$ and $\rho_m^{[P]}$, for a total of 5 months (i.e. one less than the total number of time periods since the first among six periods is conditioned upon during the estimation), and estimates for $\beta_p^{[A]}$ where p ranges from 1 to 9, and for $\beta_q^{[P]}$ where q ranges from 1 to 5.

### 4.2.1 Results on networks

For the network structure variables, as shown from the results in Tables 2(a) and 2(b), we observe that the estimate for the out-degree of the users is significantly negative (-9.536;

$p$<0.01). Since, the evaluation function can be thought of as a measure of the "fitness" or "attractiveness" of the state of the network, this estimate indicates that users in our network show a lower propensity over time to establish new social connections. This can be attributed to the cost of forming social connections or constrained resources (Dunbar 1992; Phan and Airoldi 2015). Further, we observe that the estimate for network transitivity is positive (0.109; $p$<0.01). This indicates that there is an increased drive towards network closure in our observed network. For instance, if users $i$ and $j$ are friends, and users $j$ and $h$ are friends as well, then the user $i$ has a stronger motivation to befriend user $h$ over any other user in the network, as this increases the overall attractiveness of the new network state for $i$. We also find strong evidence for friendship formation among those with a similar level of posting behavior (0.127; $p$<0.01). Thus, the more active posters prefer to befriend other active posters, while the less active posters prefer other less active ones. Interestingly, none of the other covariates were found to contribute to homophilous friendship formation.

### 4.2.2 Results on behavior

Among the behavior variables, we observe that the estimate for the linear tendency parameter is significantly negative (-0.196; $p$<0.01). As mentioned earlier, the tendency effect represents a drive towards high posting volume. A zero value on this parameter indicates user's preference for the average posting volume. Since we obtain a negative estimate on this parameter, it indicates that as time goes by, users prefer to post less. We also find strong evidence of peer influence among the students, with a significantly negative parameter for the influence effect (-2.995; $p$<0.01). This implies that individuals tend to correct their posting behavior over time in a direction away from their peers. This could be a result of free-riding behavior in case the peers are contributing more, or could also be representative of an increased drive to behave in a non-conformist manner (e.g. "*If everyone else is posting more, I should do something different*").

While it is hard to uncover the specific reasons for the peer effects we find, interpreting the parameters for homophily and peer influence together leads to an increased understanding of the interplay between friendship formation and content production behavior in online networks. Taken together, the two parameters suggest that while students prefer to befriend other students who are similar to themselves in posting behavior, they tend to move apart over time after becoming friends. Thus, behavioral similarity could play the role of a facilitator during the early days of friendship formation, but act as a deterrent in the longer run. We contend that this insight is not only theoretically important to uncover but has very strong practical implications as well, which we shall discuss in Section 5.

#### 4.2.3 Results on behavioral dependency of homophily and peer influence

In addition to the above, we also find strong evidence for the behavioral dependency of homophily and peer influence. Tables 3(a) and 3(b) illustrate the estimation results for rate parameters $\rho_m^{[A]}$ and $\rho_m^{[P]}$, for periods 2 to 6 (i.e. the first among six periods is conditioned upon during the estimation), and estimates for $\beta_p^{[A]}$ where $p$ ranges from 1 to 11 and for $\beta_q^{[P]}$ where $q$ ranges from 1 to 7. The results from the estimation show that the users in our sample demonstrate varying propensities to create homophilous relationships and varying susceptibility to peer influence, depending on the current state of their posting behavior. Specifically, compared to moderately active posters (MoAP), most active posters (MAP) were less likely to form friendships with other MAPs (-0.375; $p$<0.01), while least active posters (LAP) were more likely to form friendships with other LAPs (0.293; $p<0.01$). Further, compared to MoAPs, both MAPs and LAPs were found to be more susceptible to peer influence. However, while MAPs showed positive influence (i.e. converge in behavior with peers) (1.183; $p<0.01$), the LAPs showed negative influence (i.e. diverge in behavior from peers) (-6.437; $p<0.01$).

| Network Parameters | Estimate |
|---|---|
| Friendship rate (Period 1) | 7.767*** (0.100) |
| Friendship rate (Period 2) | 6.267*** (0.088) |
| Friendship rate (Period 3) | 3.930*** (0.082) |
| Friendship rate (Period 4) | 4.547*** (0.075) |
| Friendship rate (Period 5) | 5.353*** (0.084) |
| Out-Degree | -9.536*** (0.011) |
| Transitivity | 0.109*** (0.001) |
| Gender homophily | 0.068 (0.057) |
| Gender on Degree | 0.031 (0.020) |
| Age homophily | 0.024 (0.034) |
| Age on degree | 0.011 (0.009) |
| Tenure homophily | 0.003 (0.022) |
| Tenure on degree | -0.032** (0.016) |
| **Posting homophily** | **0.127*** (0.034)** |

| Behavior Parameters | Estimate* |
|---|---|
| Posting rate (Period 1) | 4.337*** (0.145) |
| Posting rate (Period 2) | 3.889*** (0.143) |
| Posting rate (Period 3) | 5.229*** (0.205) |
| Posting rate (Period 4) | 4.995*** (0.195) |
| Posting rate (Period 5) | 3.681*** (0.119) |
| Posting Tendency (Linear Shape) | -0.196*** (0.007) |
| **Influence** | **-2.995*** (0.134)** |
| Gender on Posting | 0.007 (0.012) |
| Age on Posting | 0.003 (0.006) |
| Tenure on Posting | 0.003 (0.010) |

*** <0.01, ** <0.05, *<0.1

**Tables 2(a) and 2(b). Estimation results for network and behavior effects**

| **Network Parameters** | **Estimate** | **Behavior Parameters** | **Estimate*** |
|---|---|---|---|
| Friendship rate (Period 1) | 7.750*** (0.100) | Posting rate (Period 1) | 4.948*** (0.274) |
| Friendship rate (Period 2) | 6.368*** (0.089) | Posting rate (Period 2) | 4.137*** (0.166) |
| Friendship rate (Period 3) | 3.997*** (0.075) | Posting rate (Period 3) | 5.477*** (0.317) |
| Friendship rate (Period 4) | 4.543*** (0.075) | Posting rate (Period 4) | 5.879*** (0.255) |
| Friendship rate (Period 5) | 5.431*** (0.084) | Posting rate (Period 5) | 4.583*** (0.252) |
| Out-Degree | -9.562*** (0.011) | Posting Tendency (Linear Shape) | -0.182*** (0.011) |
| Transitivity | 0.107*** (0.001) | **Influence** | **-2.951*** (0.268)** |
| Gender homophily | 0.076 (0.058) | **High Posters Influence** | **1.183*** (0.096)** |
| Gender on Degree | 0.023 (0.019) | **Low Posters Influence** | **-6.437*** (0.763)** |
| Age homophily | 0.023 (0.034) | Gender on Posting | 0.009 (0.012) |
| Age on degree | 0.013 (0.009) | Age on Posting | 0.004 (0.005) |
| Tenure homophily | -0.001 (0.022) | Tenure on Posting | 0.004 (0.010) |
| Tenure on degree | -0.019 (0.016) | | |
| **Posting homophily** | **0.112*** (0.042)** | | |
| **High Posters Homophily** | **- 0.375*** (0.021)** | | |
| **Low Posters Homophily** | **0.293*** (0.023)** | | |

*** <0.01 ** <0.05 *<0.1

**Tables 3(a) and 3(b). Estimation results with behavioral dependency**

## 4.3 Comparative analysis of baseline modeling approaches

In this section, we present results from two baseline approaches. The first baseline approach models online content production using a fixed effect panel linear regression model and a fixed effect Poisson regression model, as illustrated by the model specifications (25) and (26) below. Such aggregated personal networks have been commonly used in previous studies where an individual's social network is collapsed to a fixed number of sociometric variables, like the centrality measures (Kirke 2004; Yoganarasimhan 2012). These measures are then used as regressors, together with individual-level attributes, in a linear model to explain outcomes of individual-level behavior. This approach, however, ignores both homophilous friendship formation, as well as the continuous-time evolution of the network itself. The sociometric variable included is the out-degree $D_{it-1}$, which denotes the total number of friends added by the user $i$ in time period *t-1* on the SNS

$$p_{it} = \gamma Z_{it-1} + \kappa_i + \tau_t + \epsilon_{it} \tag{25}$$

for ordinary least square linear regression, or,

$$\log p_{it} = \beta Z_{it-1} + \kappa_i + \tau_t + \epsilon_{it}$$

for Poisson regression, where,

$$Z_{it-1} = \{1, D_{it-1}, P_{jt-1}, Age_i, Gender_i, SNS\ tenure_i\}' \tag{26}$$

$\kappa_i = \{\kappa_1, \kappa_2, \dots, \kappa_n\}$, for *n* individuals,

$\tau_t = \{\tau_1, \tau_2, \dots, \tau_t\}$, for a total of *t* months,

$\epsilon_{it} = \{\epsilon_{11}, \epsilon_{12}, \dots, \epsilon_{nt}\}$, and

$\gamma = \{\gamma_0, \gamma_1, \dots, \gamma_5\}$

In the above model specifications, the coefficient $\beta_2$ provides an estimate of peer influence based on the posting behavior of the peers of a user $i$. We also control for the user $i$'s biological age ($Age_i$), gender ($Gender_i$), as well as social network age ($SNS\ Tenure_i$), which is the number of days spent by the user on the SNS at the time of recording the data. The descriptive statistics for the variables were provided earlier in Table 1.

The estimation results are illustrated in Table 4 below. The results show that using a discrete-time aggregated network approach such as this leads us to believe that the peer's posting behavior has a weakly positive effect on the individual's posting behavior in the subsequent time period, after controlling for other covariates. However, as mentioned earlier, this method ignores any selection bias in friendship formation caused due to homophily and thus provides biased estimates of peer influence. The results from our co-evolution model from the previous section show that the effect of peer influence is actually the reverse (i.e. significantly negative), once we factor in homophilous friend selection into our model.

A second baseline model specifies and estimates the co-evolution of the network and behavior, but ignores both homophily based on dynamic content production, and the effect of peer's content posting behavior on the individual. Thus, we estimated a model that relies only on homophily based on stable attributes like age, gender etc., and ignores any role played by dynamic behaviors like content postings. The result from this model is illustrated in Table 5. The results from this model are consistent with our earlier results and reaffirm our belief that the students in our sample are not establishing friendships based on similarities in age, gender, or SNS tenure. Rather, they are forming new ties based on similarities in content posting behavior. Moreover, the results from this model prove that the students' content posting behavior is not influenced by their personal attributes such as age, gender or SNS tenure, but are instead influenced largely by the posting behavior of their peers, as was illustrated in Tables 2 and 3.

| **Variables** | **(Random-effects Panel Linear Regression) Posts ($P_{it}$)** | **(Fixed-effects Panel Linear Regression) Posts ($P_{it}$)** | **(Fixed-effects Poisson Regression) Posts ($P_{it}$)** |
|---|---|---|---|
| $D_{it-1}$ | | -6.381*** (0.814) | -0.007*** (0.0001) |
| $(\sum_j P_{jt-1})$ | | 0.0001** (0.0001) | 0.0000002*** (0.00000001) |
| $Age_i$ | -77.831*** (8.245) | (omitted) | -0.789*** (0.043) |
| $SNS\ tenure_i$ | 0.150*** (0.033) | (omitted) | 0.001*** (0.0002) |
| $Gender_i$ (=1) | 15.072 (58.176) | (omitted) | -0.076 (0.289) |
| $Gender_i$ (=2) | -9.263 (58.144) | (omitted) | -0.196 (0.288) |
| $Gender_i$ (=4) | -120.111 (368.536) | (omitted) | -22.485 (3763.516) |
| Time dummies | Present | Present | Present |
| Sample size | 2030 | 2012 | 2012 |
| R-squared | 0.040 | 0.052 | |

*** <0.01 ** <0.05 *<0.1

**Table 4. Results from discrete-time aggregated network models**

| Network Parameters | Estimate | Behavior Parameters | Estimate |
|---|---|---|---|
| Friendship rate (Period 1) | 7.558*** (0.098) | Posting rate (Period 1) | 3.064*** (0.100) |
| Friendship rate (Period 2) | 6.226*** (0.097) | Posting rate (Period 2) | 2.804*** (0.114) |
| Friendship rate (Period 3) | 3.977*** (0.076) | Posting rate (Period 3) | 3.694*** (0.149) |
| Friendship rate (Period 4) | 4.727*** (0.080) | Posting rate (Period 4) | 3.510*** (0.117) |
| Friendship rate (Period 5) | 5.608*** (0.087) | Posting rate (Period 5) | 2.679*** (0.100) |
| Out-Degree | -9.070*** (0.010) | Posting Tendency (Linear Shape) | -0.116*** (0.008) |
| Transitivity | 0.056*** (0.001) | Gender on Posting | -0.003 (0.013) |
| Gender homophily | 0.007 (0.056) | Age on Posting | 0.006 (0.012) |
| Gender on Degree | 0.007 (0.019) | Tenure on Posting | -0.002 (0.011) |
| Age homophily | -0.006 (0.023) | | |
| Age on degree | 0.006 (0.016) | | |
| Tenure homophily | -0.007 (0.021) | | |
| Tenure on degree | -0.002 (0.016) | | |

*** <0.01, ** <0.05, *<0.1

**Table 5. Estimation results for network and behavior based on covariates alone**

### 4.4 Sensitivity to latent homophily

While our analysis conditions on observable behavioral (e.g. posting) and individual-level covariates, (e.g. age and gender), there is a possibility that the network formation might be driven by homophily based on latent factors, such as personality traits and similarity in tastes or preferences. The presence of such latent homophily has been cited as an important confound in the estimation of social influence (Shalizi and Thomas 2011). We look to test the sensitivity of our modeling approach to the presence of such latent homophily using a latent space modeling approach, similar to what has been described in Davin et al. (2014). Latent space models are well known in social networks literature and have been traditionally employed in identifying and visualizing communities within networks. For our analysis, we use 2-dimensional latent space positions as proxy variables to control for potential latent homophily. The intuition behind this approach is that if two actors are close to each other in a latent social space, then this similarity is driven by both observed as well as unobserved factors. Thus, adding latent space coordinates as model covariates would serve to reduce the bias associated with influence estimate by controlling for some latent homophily. There have been some prior work that have used latent space models to address similar questions in economics and marketing (Ansari et al. 2011; Braun and Bonfrer 2011). A summary of how the latent space models for our current context were specified and estimated has been illustrated in Appendix 6.

We estimate the rate and evaluation functions from the co-evolution model as specified in Sections 3.1 and 3.2.1 earlier, using the latent space positions as coordinates, and present the results in Table 6. We find that the results for both homophily based on posting behavior as well as peer influence are consistent with our previous results. As expected, after controlling for homophily based on latent space coordinates, the estimate for posting homophily (0.104; $p<0.01$) reduces in strength, but continues to be statistically significant. This shows that there does exist evidence of homophily based on latent factors beyond the observable factors of age, gender and SNS tenure. However, our proposed effect of posting homophily exists even after controlling for possible latent confounders. Similarly, the estimate for peer influence is weaker (-0.015; $p<0.01$) than our earlier models that do not account for latent homophily. In summary, we leverage latent space positions of actors in our network to account for possible latent homophily, and show that our results for homophily and peer-influence based on posting behavior are valid even after controlling for these latent positions.

| Network Parameters | Estimate | Behavior Parameters | Estimate |
|---|---|---|---|
| Friendship rate (Period 1) | 7.529*** (0.098) | Posting rate (Period 1) | 3.496*** (0.182) |
| Friendship rate (Period 2) | 6.204*** (0.141) | Posting rate (Period 2) | 3.590*** (0.132) |
| Friendship rate (Period 3) | 3.949*** (0.075) | Posting rate (Period 3) | 4.223*** (0.114) |
| Friendship rate (Period 4) | 4.586*** (0.085) | Posting rate (Period 4) | 4.603*** (0.122) |
| Friendship rate (Period 5) | 5.487*** (0.091) | Posting rate (Period 5) | 3.104*** (0.104) |
| Out-Degree | -9.913*** (0.014) | Posting Tendency (Linear Shape) | -0.191*** (0.008) |
| Transitivity | 0.098*** (0.001) | Influence | -0.015*** (0.001) |
| Gender homophily | 0.048 (0.077) | Gender on Posting | 0.007 (0.014) |
| Gender on Degree | 0.011 (0.021) | Age on Posting | 0.008 (0.012) |
| Age homophily | 0.011 (0.032) | Tenure on Posting | -0.002 (0.012) |
| Age on degree | 0.005 (0.024) | *** <0.01, ** <0.05, *<0.1 | |
| Tenure homophily | -0.010 (0.030) | | |
| Tenure on degree | -0.012 (0.018) | | |
| Posting homophily | 0.104*** (0.047) | | |
| Latent Pos. (X) homophily | 1.030*** (0.126) | | |
| Latent Pos. (Y) homophily | 1.110*** (0.112) | | |

**Table 6. Latent homophily corrected estimation results for network and behavior**

# 5 Discussion and Conclusion

In the current study, we develop and estimate a model for analyzing the co-evolution of content production and social network structure using real world data from a large social network site. Our results demonstrate the role of social network structure and user-characteristics in influencing content production on SNS. We adopt an actor-driven and co-evolution based MCMC modeling approach to jointly estimate the evolution of the user's social network and posting behavior. We contend that this approach is more statistically disciplined than several previous methods, which tend to violate some key assumptions of network-based modeling. Furthermore, we depart from previous instances of the actor-driven models whose applicability is restricted to stable dichotomous behaviors, like smoking, and substance abuse. In the current study, we adapt the co-evolution model to a dynamic behavior (i.e. online public posts) which often changes rapidly over successive time periods. We avoid convergence related difficulties with MCMC estimations of such continuous behavioral variables by discretizing our behavioral variable into several quantiles to represent the intensity of behavior. We contend that by using this quantile-based binning strategy, we are able to achieve high convergence in estimations without much loss of information. Furthermore, we account for homophilous friend selection based on unobserved covariates i.e. latent homophily, by including latent space positions of the actors as covariates in our model. The results from our analyses uncover important insights about how users make friends on SNS, and how the network, in turn, influences their content production behavior. Specifically, we show that users are more likely to make friends with users who show a similar level of posting behavior, as observed by the number of public posts. However, this homophilous behavior is short-lived and the users are found to diverge in their content production rates from their peers over time. Furthermore, our analyses shows that the propensity to form friendships based on homophily, and the susceptibility to peer influence after forming the friendships, are dependent on the current state of the behavior. Thus, users who are very active contributors on SNSs show very different peer effects as compared to users who are less active on the SNS.

## 5.1 Theoretical implications

Using our co-evolution perspective, we address the following two theoretical gaps in the existing research on the evolution of online social networks and social behavior.

First, we show that homophilous peer selection and peer-influence might have varying strengths depending on the stage of network evolution. We find strong evidence of selection bias on the basis of homophily in content production, i.e., the students make friends with others who are similar in their content production behavior. Once they become friends, however, our findings show that they exhibit a negative influence effect. This means that the students actively try to distinguish themselves from their friends in terms of their content production behavior. This is an interesting phenomenon, which demonstrates that dynamic behaviors such as content production can influence network evolution in competing ways.

Second, we uncover a behavioral dependency of these network effects, such that homophilous selection and peer-influence increase or decrease in strength as a function of the current magnitude of an individual's behavior. We find that students who are very active content producers (i.e. MAP users) are qualitatively very different from students who are highly inactive producers (i.e. LAP users) and students who are moderately active content producers (i.e. MoAP users). These three groups of students displayed different degrees of inclination towards homophilous peer selection and different degrees of susceptibility towards peer-influence. Taken together, these results reveal an interesting pattern of how online social networks co-evolve with the content produced on these platforms.

## 5.2 Practical contributions

Understanding the nature of peer effects on SNS has clear practical implications for several stakeholders. Firstly, and most importantly, we offer a framework within which online user contributions can be studied as a function of the underlying network. While it is common for researchers and practitioners to use predictive and explanatory models of social media content production, they often tend to ignore the underlying social network that connects the content producers. We offer a robust statistical model to help explain content production while being conscious of the evolution in the underlying network structure. This would help platform owners and marketers derive more reliable insights about their users.

Secondly, our results provide intelligence to marketers to identify and better target valuable users on SNS. Understanding what drives content production on online platforms, and the impact of peers on the user’s propensity to produce content is key to devising better strategies to enable and sustain content production on the platform. Moreover, by understanding how friendships are created and altered over time, platforms like Facebook and Twitter can help improve friend recommendations and personalized content through customized "newsfeeds". Specifically, our results suggest that it might not be a good idea to recommend heavy content posters as friends to other heavy posters, as such friendships tend to be detrimental to the content production of either of the friends, i.e., high posters prefer other high posters in making friends, but reduce their posting rate over time after the friendship is created. Moreover, we also show that this tendency to alter behavior in response to peers is strongest for heavy posters and weakest for low posters. Thus, the findings from this study can guide platform owners on better managing their active content producers.

Thirdly, our model also allows for predictive analysis of posting behavior on these platforms, such that managers and researchers can effectively seed content, and forecast the diffusion of this content through social networks. Such predictive models for user behavior on dynamic networks can be invaluable not just to the platform owners, but also to advertisers and third-party marketers who wish to leverage social media for their own businesses. Thus, we believe that the specific findings from our study and the methodology in general can increase content-creation and user retention in such SNS platforms.

### 5.3 Limitations and future work

As an initial attempt to model and analyze the co-evolution of network structure and user behavior in online social networks, this study is prone to several limitations that offer opportunities for future research. Firstly, and as mentioned earlier, the current paper focuses on providing a statistically sound method to uncover the dynamic peer effects in a university social network. However, additional analyses are required to further separate out the specific rationale behind why individuals show such effects. Secondly, our current modeling approach requires computational resources to simulate the networks in each stage of the estimation procedure. This might be a concern for extremely large networks of users, and networks with high sparsity. In such cases, we might have to resort to bootstrapping approaches which introduce concerns about network-based sampling, a non-trivial area of active research in its own right. Our model imposes a standard Markovian assumption on the data, which is reasonable in most cases. However, this assumption implies that there are no external factors that might influence the social network or the user behavior e.g. shocks in the physical world. Even though we have controlled for common covariates that have been used in recent social network studies (e.g. age, gender and experience), and accounted for the possibility of unobserved confounds that might play a role, there is a possibility of external events (e.g. term breaks) to affect the propensity to make friends and the subsequent behavioral influence. Lastly, we consider all friendships to be bi-directional or symmetric ties. While this is not a limitation in the present study, it could be useful to identify the directionality of friendship i.e. separate out in-degree from out-degree. While in-degree can be considered to be a measure of popularity, out-degree provides a better indication of SNS activity. Thus, by separating out the two effects, we will be able to investigate more complex social constructs in future studies.

# Appendix 1: Stochastic Approximation

In our study, we employ a Method of Moments (MoM) estimation procedure for the model specified in Sec. 3.1.3 (Bowman and Shenton 1985). The MoM estimator for our data (A,P) and parameter sets $\theta^{[A]}$and $\theta^{[P]}$ is based on a set of network and behavioral statistics $S_t^{[A]}$ and $S_t^{[P]}$, and is defined as the parameter value setfor which the following conditions are satisfied.

$$E_{\theta^{[A]}}\left(S^{[A]}\right) = s^{[A]}(a, p) \tag{i}$$

$$E_{\theta^{[P]}}\left(S^{[P]}\right) = s^{[P]}(a, p) \tag{ii}$$

i.e., the expected values and the observed values of the statistics are the same.

The choice of network and behavior statistics have been discussed in Sec. 3.2. In the general case, conditional expectations from the moment equations (i and ii) cannot be computed explicitly. Thus, we use a stochastic approximation method (Robbins and Monro 1951) to solve these moment equations. The method used to solve Eqs. i and ii involves iteratively generating a parameter sequence $\hat{\theta}$ according to the following iteration steps.

$$\hat{\theta}_{t+1}^{[A]} = \hat{\theta}_t^{[A]} - \sigma_t D_0^{-1}(S_t^{[A]} - s^{[A]}) \tag{iii}$$

$$\hat{\theta}_{t+1}^{[P]} = \hat{\theta}_t^{[P]} - \sigma_t D_0^{-1}(S_t^{[P]} - s^{[P]}) \tag{iv}$$

where, $S_t^{[A]}$ and $S_t^{[P]}$ are generated according to the distributions defined by $\hat{\theta}_t^{[A]}$ and $\hat{\theta}_t^{[P]}$ respectively. The step size $\sigma_t$ needs to be a sequence that converges to zero. The sequence $\sigma_t = \frac{a}{b + t}$ for any two integers a and b satisfies this constraint. $D_0^{-1}$ is an identity matrix.

Snijders (2001) shows that the convergence properties of this algorithm hold asymptotically for t ->∞ (Polyak 1990; Ruppert 1988; Yin 1991).

# Appendix 2

*(a) :Descriptive summary for social network data*

| | Time Period | | | | | |
|---|---|---|---|---|---|---|
| **Observation** | **1** | **2** | **3** | **4** | **5** | **6** |
| **Density** | 0.025 | 0.027 | 0.028 | 0.029 | 0.030 | 0.031 |
| **Average Degree *** | 63.276 | 67.165 | 70.377 | 72.42 | 74.844 | 77.747 |
| **Number of Ties** | 79317 | 84191 | 88217 | 90778 | 93817 | 97456 |
| **Missing Fraction** | 0 | 0 | 0 | 0 | 0 | 0 |

*** Average degree across all periods = 70.971**

*(b) :Social network evolution summary*

| | Change in Ties | | | | | |
|---|---|---|---|---|---|---|
| **Period** | **0 => 0** | **0 => 1** | **1 => 0** | **1 => 1** | **Jaccard *** | **Missing** |
| **1==>2** | 3057080 | 4874 | 0 | 79317 | 0.942 | 0 (0%) |
| **2==>3** | 3053054 | 4026 | 0 | 84191 | 0.954 | 0 (0%) |
| **3==>4** | 3050493 | 2561 | 0 | 88217 | 0.972 | 0 (0%) |
| **4==>5** | 3047454 | 3039 | 0 | 90778 | 0.968 | 0 (0%) |
| **5==>6** | 3043815 | 3639 | 0 | 93817 | 0.963 | 0 (0%) |

* Jaccard Index = $\frac{N_{11}}{N_{01}+N_{10}+N_{11}}$ , where $N_{hk}$ is the number of tie variables with value $h$ in one wave, or observation from our dataset, and the value $k$ in the next wave.

# Appendix 3

*2(a) Descriptive summary for behavior data*

| | Time Period | | | | | |
|---|---|---|---|---|---|---|
| **Posting quantile** | **1** | **2** | **3** | **4** | **5** | **6** |
| **1 (lowest)** | 630 | 763 | 787 | 644 | 806 | 774 |
| **2** | 945 | 1027 | 1019 | 903 | 1015 | 1009 |
| **3** | 364 | 317 | 324 | 326 | 325 | 336 |
| **4** | 193 | 177 | 157 | 223 | 147 | 147 |
| **5** | 122 | 86 | 76 | 118 | 85 | 88 |
| **6** | 77 | 47 | 41 | 84 | 53 | 54 |
| **7** | 33 | 24 | 37 | 50 | 18 | 30 |
| **8 (highest)** | 26 | 18 | 19 | 45 | 18 | 20 |

*Note: The figures in the cells indicate the number of users who have posted in that time period. Row 1 indicates the total number of first-quantile posters (i.e. low posters) in each of the 6 time periods. Similarly, Column 1 indicates the number of posters in each of the 8 posting quantiles for the first time period.*

*2(b) Behavior evolution summary*

| | Number of users | | | |
|---|---|---|---|---|
| **Period** | **Decrease Posting Behavior** | **Increase Posting Behavior** | **Constant** | **Missing** |
| **1 => 2** | 1009 | 427 | 1071 | 0 |
| **2 => 3** | 674 | 653 | 1180 | 0 |
| **3 => 4** | 378 | 1057 | 1072 | 0 |
| **4 => 5** | 1066 | 367 | 1074 | 0 |
| **5 => 6** | 625 | 711 | 1171 | 0 |

# Appendix 4

Convergence Assessment for Network Variables

| Network Variables | Observed Value for Target Statistics | Av. Deviation of simulated statistic from target statistic (SD Deviation) |
|---|---|---|
| Friendship rate (Period 1) | 9748.000 | -370.044 (139.914) |
| Friendship rate (Period 2) | 8052.000 | -141.393 (126.777) |
| Friendship rate (Period 3) | 5122.000 | 24.940 (97.883) |
| Friendship rate (Period 4) | 6078.000 | 63.709 (108.249) |
| Friendship rate (Period 5) | 7278.000 | 254.976 (120.326) |
| Out-Degree | 454459.000 | -83.906 (131.380) |
| Transitivity (No. of triads) | 4445064.000 | -2548.047 (3834.512) |
| Gender on Degree | 11547.242 | -76.155 (108.814) |
| Gender homophily | -2109.296 | -37.071 (37.992) |
| Age on degree | -29955.212 | -209.697 (276.408) |
| Age homophily | -6381.205 | -45.223 (68.114) |
| Tenure on degree | 19104.656 | 186.366 (157.918) |
| Tenure homophily | -1968.574 | 12.749 (98.977) |
| Posting homophily | 22210.253 | -199.796 (80.153) |

# Appendix 5

Convergence Assessment for Behavior Variables

| **Behavior Variables** | **Observed Value for Target Statistics** | **Av. Deviation of simulated statistic from target statistic (SD Deviation)** |
|---|---|---|
| Posting rate (Period 1) | 2150.000 | -33.206 (47.848) |
| Posting rate (Period 2) | 1663.000 | -25.970 (46.780) |
| Posting rate (Period 3) | 2359.000 | -89.171 (51.167) |
| Posting rate (Period 4) | 2310.000 | -67.194 (50.710) |
| Posting rate (Period 5) | 1614.000 | -77.020 (46.518) |
| Posting Tendency (Linear Shape) | 1801.000 | -5.009 (129.923) |
| Influence | 1798.000 | 7.562 (15.943) |
| Gender on Posting | 1825.000 | 3.782 (78.867) |
| Age on Posting | 1755.000 | -8.259 (220.323) |
| Tenure on Posting | 1900.000 | -8.098 (119.231) |

# Appendix 6: Latent Space Model

Following past work on statistical network models on exponential random graph models (Frank and Strauss 1986; Wasserman and Pattison 1996), the homogenous monadic Markov model (Frank and Strauss 1986), the stochastic and mixed membership block-models (Airoldi et al. 2008; Wang and Wong 1987), and the latent class membership models (Nowicki and Snijders 2001), Hoff et al. proposed a statistical approach to represent network actors as points on a latent social space (Hoff et al. 2002). The actors' positions on this Euclidean space are a result of the actors' observed as well as unobserved characteristics, and hence, the distance between these points is reflective of any underlying latent homophily based on these unobserved factors. The latent space model emphasizes conditional independence of the relational ties such that, conditional on the positions of the actors in the latent space, the probabilities of the tie formation are independent of each other. The latent space model is specified as follows:

$$\Pr(A|Z,X,\theta) = \prod_{i\neq j} P(a_{ij}|z_i,z_j,x_{i,j},\theta)$$

where, $x_{i,j}$ is a vector of the observed covariates comprising similarity based on age, gender and SNS tenure, and $z_i$ captures the latent space positions of actor $i$. The latent position vector Z and the parameter set $\theta$ are both estimated from the model. Now, a convenient specification for the tie-formation probability $P(a_{ij}|z_i,z_j,x_{i,j},\theta)$ is the logistic regression model as follows:

$$\eta_{ij} = \log odds\,(y_{ij} = 1\,|z_i,z_j,x_{i,j},\alpha,\beta) = \alpha + \beta' x_{i,j} - |z_i - z_j|$$

We follow (Hoff et al. 2002) and assume that the $z_i's$ are independent draws from a spherical multivariate normal distribution as follows:

$$z_1, z_2 \dots z_N \sim MVN_k(0, \sigma_Z^2 I_k)$$

where, N is the sample size, k is the dimension of the latent space,

The log-likelihood for the above latent space model is then constructed as follows:

$$log\ Pr(A|\eta) = \sum_{i \neq j}\{\eta_{ij} a_{ij} - \log(1 + e^{\eta_{ij}})\}$$

where, $\eta_{ij}$ is the log odds of tie formation and given as $\alpha + \beta' x_{i,j} - |z_i - z_j|$. As is clear from the formulation of the log likelihood, the computation of this function requires a sum over N(N-1) terms, which leads to a run-time complexity of O($N^2$). This makes the direct MLE estimation infeasible for large-sample networks datasets, such as ours. We perform the likelihood based inference by following an approximation strategy proposed in (Raftery et al. 2012) which reduces the computational cost from O($N^2$) to O(N). The approximation uses a case-controlled approach as popularized by Breslow (1996) and Breslow et al. (1980) but with a stratified sampler, to represent the likelihood as a sum of case likelihood (for $a_{ij} = 1$) and control likelihood (for $a_{ij} = 0$). We then estimate the approximate likelihood using a MCMC estimator.